\documentclass[12pt]{iopart}
\usepackage{graphicx}  
\begin{document}

\title[Inclusion and validation of electronic stopping in the open source
LAMMPS code] {Inclusion and validation of electronic stopping in the open
source LAMMPS code}

\author{H. Hemani$^1$, A. Majalee$^1$, U. Bhardwaj$^1$, A. Arya$^{2,3}$,
K. Nordlund$^4$, M. Warrier$^{1,3}$}

\address{$^1$ Computational Analysis Division, BARC, Visakhapatnam, Andhra
         Pradesh – 531011, India}
\address{$^2$ Glass \& Advanced Materials Division, BARC, Trombay, Mumbai
         – 400085, India}
\address{$^3$ Homi Bhabha National Institute, Anushaktinagar, Mumbai - 400094,
         India}
\address{$^4$ Accelerator Laboratory, Pietari Kalmin katu 2 P.O.Box 43 Helsinki,
            Finland}
\ead{harsh@barc.gov.in}

\begin{abstract}
Electronic stopping (ES) of energetic atoms is not taken care of by the 
interatomic potentials used in molecular dynamics (MD) simulations when 
simulating collision cascades. The Lindhard-Scharff (LS) formula for electronic 
stopping is therefore included as a drag term for energetic atoms in the open 
source large scale atomic molecular massively parallel simulator (LAMMPS) code. 
In order to validate the ES implementation, MD simulations of collision cascades 
at primary knock-on atom (PKA) energies of 5, 10 and 20 keV are carried out in 
W and Fe in 100 random directions. The total ES losses from the MD simulations 
show energy straggling due to the stochastic nature of the phenomena. The
losses due to ES are compared with that predicted by the
Norgett-Robinson-Torrens (NRT) model to validate our implementation. It is seen
that the root mean square deviation of ES losses from 
the MD implementation is around 10 \% for both W and Fe compared to the NRT 
model. The velocity threshold above which electronic stopping is important is
explored. The effect of ES on the number of defects in collision cascades is
presented for Fe and W.
\end{abstract}

\vspace{2pc}
\noindent{\it Keywords}: Electronic Stopping Collision Cascades, Molecular 
Dynamics, LAMMPS
%\submitto{\MSMSE}
%\maketitle

\section{Introduction}

Energetic ions impinging on a target are slowed down by interactions with the 
nuclei of atoms and with the bound electrons of the target. The interactions 
with the nuclei depends on the repulsive potential between the atoms and is 
called nuclear stopping (NS). The interactions with the bound electrons of the 
material, and with the free electron gas if the material is a metal, is 
referred to as electronic stopping (ES). ES dominates NS at energies of a few 
tens of keV, the exact value depending on the material under consideration. The 
interatomic potentials used in classical MD simulations are usually obtained by 
fitting equilibrium properties of materials. This does not include the 
inelastic collisions between a fast particle and electrons in the system – i.e. 
electronic stopping. Therefore ES has to be accounted for separately in MD 
simulations of collision cascades.

Nuclear stopping depends on the repulsive potential between the energetic atom 
and the atoms of the material and is described by the semi-empirical Ziegler, 
Biersack and Littmark (ZBL) potential \cite{Ziegler:1525729}. The ZBL potential 
is widely used to modify interatomic potentials in molecular dynamics (MD) 
simulations of collision cascades \cite{Stoller2012293,KaiNatureComm2018}.
%!! Bring in stiffening of interatomic potentials here and comment that nuclear
% stopping is taken care of by IPs with references.

The easiest way to include the effect of ES in a molecular dynamics (MD) code 
is by damping the velocity of an atom by a viscous force:
\begin{equation}
\label{FViscous}
F_{es} = \beta v
\end{equation}
where $\beta$ is the drag co-efficient having units of mass/time, and, $v$ is
the velocity of the energetic atom [3]. The value of $\beta$ can be obtained
by several ways:
\begin{enumerate}
\item Firsov's model of mean electronic excitations in atomic collisions
      \cite{FirsovESModel},
\item Lindhard and Scharff's formula for energy dissipation of energetic ions
      \cite{PhysRev.124.128},
\item Finnis, et.al. formula taking into account the energy transfer from ions 
      to a background electron gas at a specified temperature
      \cite{PhysRevB.44.567},
\item Caro et.al's. formula including local density effects which can be
      implemented for interatomic potentials that use the local electron
      density \cite{PhysRevA.40.2287}, and
\item Nordlund, et.al., who use SRIM \cite{Ziegler:1525729} stopping power data
      for the ES \cite{PhysRevB.57.7556}.
\end{enumerate}

A different approach is taken by Duffy et.al., wherein the energy is 
transferred from an energetic ion to a background electron gas by frictional 
forces. This heats up the electron gas and subsequently raises the local 
temperature by thermal energy transfer between the electron gas and the ambient 
nuclei \cite{duffy2006including,rutherford2007effect}. This local hot-spot 
formation can have implications on the final number of defects produced and on 
in-cascade defect clustering. Mason et.al., have modeled the energy loss from 
ions to electrons using Ehrenfest dynamics \cite{le_Page_2008} with a tight 
binding Hamiltonian wherein the electrons are not assumed to be in their ground 
state. They also, like Duffy et.al., treat the subsequent heat transfer from 
electrons to the ambient nuclei using a Langevin thermostat. They conclude 
that, ``A simple homogeneous viscous damping coefficient is adequate to model 
the energy transfer rate between ions and electrons, but the accuracy can be 
improved by making the damping constant dependent on the electron density and 
by computing the probabilities of specific transitions between bands''
\cite{RACE201233}.

LAMMPS \cite{PLIMPTON19951} is an open source software capable of massively 
parallel atomistic simulations. We use it widely in our radiation damage 
studies, not only to simulate colision cascades 
\cite{warrier2015statistical,warrier2016multi}, but also to simulate defect 
diffusion \cite{BHARDWAJ2016263,bukkuru2017identifying}. It has several package 
extensions amongst which the two temperature model (TTM) by Duffy et.al., is 
also implemented. Using the TTM, involves inputting the electron thermal 
conductivity and the electron-ion interaction co-efficient which is not 
available for many materials of our interest. More recently, electronic 
stopping in LAMMPS has been introduced as a friction term with a low energy 
cut-off. It however needs the stopping powers to be input by the user. We 
calculate the stopping power within our module using the Lindhard-Scharff model 
to obtain a friction co-efficient and implement damping of the velocities of 
energetic atoms above a specified cut-off energy in LAMMPS. We validate our 
implementation by comparing the total energy lost by ES with the electronic 
loss expected from the standard NRT model \cite{NORGETT197550}.

This paper aims to describe the methods we have used and the validations 
carried out to include ES in LAMMPS. The next section describes the inclusion 
of ES as a friction term in MD simulations. Section.3 describes 
the MD runs with ES included to validate the included code for ES. 
Section.4 presents the results. The pros and cons of the three 
implementations of electronic stopping in LAMMPS mentioned in the previous 
paragraph is discussed at the end of section.4. Finally we present the 
conclusions.

\section{Including Electronic Stopping in LAMMPS}
\label{IncESinLAMMPS}
ES is implemented in LAMMPS as a “fix” command. It is invoked as follows:\\
\begin{verbatim}
fix       fix-id group-id esfriction Z1 Z2 A1 A2 neval Epka vel-cutoff
\end{verbatim}
where, $Z1$ and $Z2$ are the atomic numbers of the projectile and target atoms 
respectively and $A1$ and $A2$ are their atomic weights. The current 
implementation is for the case where the projectile and target atoms are the 
same. $neval$ is the number of valence electrons of the background target 
element, $Epka$ is the initial energy of the PKA and $vel-cutoff$ is the cutoff 
velocity below which there is no damping of velocities due to ES. In the 
following sub-sections we describe (i) the Lindhard-Scharff model for ES which 
we have implemented to obtain the damping factor $\beta$, and (ii) the total 
electron stopping losses calculation in the NRT model for validation of our 
implementation by obtaining the total energy loss due to ES.

\subsection{The Lindhard Scharff Model for Electronic Stopping}
\label{LSModelES}
The stopping force due to electronic drag by a target on an incoming atom 
having a kinetic energy E is given by
\begin{equation}
\label{ForceES}
F_{es} = \frac{dE}{dx} = n S_{es}(E)
\end{equation}
where, $S_{es}$ is the electronic stopping cross section per target atom and n 
is the number density of the target atoms. $S_{es}$ is given by
\begin{equation}
\label{LSStope}
S_{es}(E) = 8 \pi e^2 a_o Z_1^{1/6} \frac {Z_1 Z_2}{(Z_1^{2/3}+Z_2^{2/3})^{3/2}}
            \frac{v}{v_o} = \lambda v
\end{equation}
where $v$ is the velocity of the energetic atom, $Z_1$ and $Z_2$ are the atomic 
numbers of the energetic atom and of the target respectively, $a_o$ is the Bohr 
radius, $e$ is the electronic charge, $v_o$ is the Fermi velocity for the 
target atoms.

Therefore from Eqns.[1-3], we have the drag co-efficient, $\beta = n \lambda$, 
where $\lambda$ is given by
\begin{equation}
\label{LambdaExpr}
\lambda = 8 \pi e^2 a_o Z_1^{1/6} \frac {Z_1 Z_2}{(Z_1^{2/3}+Z_2^{2/3})^{3/2}
          v_o}
\end{equation}

The Fermi velocity is the velocity is given by
\begin{eqnarray}
\label{FermiVel}
v_o & = & \sqrt{\left ( \frac {2\ E_o}{m_e} \right )} \\ \nonumber
    & = & 5.9310 \times 10^5 \sqrt{E_o} \ \ \  (in\ units\ of\ m/sec)
\end{eqnarray}
where $m_e$ is the electron mass and Eo is the Fermi Energy given by: 
\begin{eqnarray}
\label{FermiEngy}
E_o & = & \frac{\hbar^2}{2m_e} (3 \pi^2 Ne_{val}*n)^{2/3}
\end{eqnarray}
where $Ne_{val}$ is the number of electrons in the outermost shell of the atom 
and $n$ is the number density of the target atoms. $n$ can easily be obtained 
from the lattice constant $c$ and the number of atoms in an unit cell $nat$ 
using $n = nat/c^3$.

From Eqns. [4-6], the drag co-efficient for any projectile – target combination 
can be obtained. 

\subsection{Total ES loss from the NRT Model}
\label{ESLossNRT}
The total loss due to ES $(Q_{es})$ for an energetic primary knock-on atom 
(PKA) in the NRT model is based on the Lindhard model. It can be obtained by 
finding out $E_{PKA} - \hat{E}$, where $\hat{E}$ is the energy available to 
generate displacements after accounting for ES losses and is approximated by
\begin{equation}
\label{EHat}
\hat{E} = \frac{E_{PKA}}{[1 + k g(\epsilon)]},
\end{equation}

where $g(\epsilon)$ is given by
\begin{equation}
\label{geps}
g(\epsilon) = 3.4008\ \epsilon^{1/6}\ +\ 0.40244\ \epsilon^{3/4}\ +\ \epsilon
\end{equation}

\begin{equation}
\label{kval}
k = 0.1337\ Z_1^{1/6}\ \left (\frac{Z_1}{A_1}\right )^{1/2}
\end{equation}

\begin{equation}
\label{epsilon}
\epsilon = \frac{A_2\ E_{PKA}}{A_1+A_2}\ \frac{a}{Z_1\ Z_2\ e^2}
\end{equation}
where $A_1$ and $A_2$ are the atomic numbers of the energetic atom and the
target atom respectively and $a$ is given by:

\begin{equation}
\label{aval}
a = \left( \frac{9\ \pi^2}{128} \right )^{1/3} \ a_o\ [Z_1^{2/3}\ +\ 
    Z_2^{2/3}]^{-1/2}
\end{equation}

Thus the total energy loss of the PKA due to ES, $Q_{es}$, is obtained from 
Eqns.[7-11].

\subsection{Fixing a Low Energy Cut-off for Electronic Stopping}
\label{LowECut}
The inter-atomic potentials used in MD simulations are good enough to simulate
atomic oscillations about a mean position and ES does not play a part in this
dynamics. We therefore make the assumption that ES needs needs to be invoked
only when an atom acquires sufficiently high energy to be displaced from its
lattice position and settle at a new interstitial / lattice position. Therefore,
the velocity corresponding to its threshold displacement energy, $E_D$, is a
good choice for the velocity cutoff for electronic stopping, $vcut_{es}$.
However, $E_D$ depends on the direction in which an atom is launched prior to
displacement by Nordlund et al., \cite{NORDLUND2006322, Juslin200775}. They use
MD simulations to obtain a direction specific threshold 
displacement energy which is a function of angles $\theta$ and $\phi$, 
$E_D(\theta,\phi)$, for each lattice direction \cite{NORDLUND2006322, 
Juslin200775}. An average threshold energy, $E_D^{avg}$, defined as the average 
of the function over all angles, can then be used as a Heaviside step function
to decide if an atom displaces or not.

A slightly different approach is described here to obtain the threshold of 
displacement. MD simulations in 1000 random directions are carried out in the 
energy range 30-500 eV to obtain a displacement energy in Fe and W 
\cite{BARCRepEDispW}. From the simulations, we obtain a direction averaged 
probability for displacement, which has a lower cutoff, $E_D^{min}$, below 
which there was no displacement and a upper cutoff, $E_D^{max}$, above which 
the probability of displacement is 1. For Fe, we obtained $E_D^{min} = 30$ eV, 
$E_D^{avg} = 60$ eV, $E_D^{max} = 90$ eV and for W we obtain $E_D^{min} = 30$ 
eV, $E_D^{avg} = 150$ eV, $E_D^{max} = 270$ eV.

MD simulations of collision cascades with our implementation of electronic 
stopping with $vcut_{es}$ corresponding to $E_D^{min}$, $E_D^{max}$ and their 
average value, say $E_D^{avg}$ have been carried out for Fe and W. Details of 
the simulations are described in the next section.

\section{Validating the Implementation of Electronic Stopping}
\label{SecValidate}
MD simulations of collision cascades were carried out at 1, 5, 10 and 20 keV in 
W and Fe to validate the implementation of ES in LAMMPS. $52 \times 52 \times 
52$ unit cells sample of W crystal was initially equilibrated using an NPT 
ensemble at 300 K temperature and 0 bar pressure. Similarly for Fe, $52 \times 
52 \times 52$ unit cells for 1 keV and 3 keV, $72 \times 72 \times 72$ unit 
cells for 5 keV, $102 \times 102 \times 102$ unit cells for 10 keV and $132 
\times 132 \times 132$ unit cells for 20 keV PKA were used and the crystal was 
equilibrated. The collision cascade simulations were then carried out by 
directing a centrally located primary knock-on atom (PKA) along 100 random 
directions at each of these energies using an NVE ensemble for 10 ps, amounting 
to a total of 400 MD simulations. The atoms in the outermost cells of the 
system were held fixed. The atoms in two unit cells lying just inside the 
outermost unit cell were subjected to an NVT ensemble at 300 K. Three sets of 
runs at each energy, for both Fe and W, were carried out corresponding to 
$vcut_{es}$ corresponding to $E_D^{min}$, $E_D^{max}$ and $E_D^{avg}$ as 
discussed in the previous section. The runs were repeated without ES to study 
the effect of ES on the number of defects produced. The number of defects as a 
function of energy for both Fe and W are calculated using the methods described 
in \cite{warrier2016multi}. The total ES losses from the LAMMPS simulations are 
compared with the total ES losses obtained from the NRT model to validate the 
\begin{verbatim} fix_esfriction \end{verbatim} code added to LAMMPS.

\section{Results and Discussions}
\label{SecResults}
\begin{figure}
  \includegraphics[width=\linewidth]{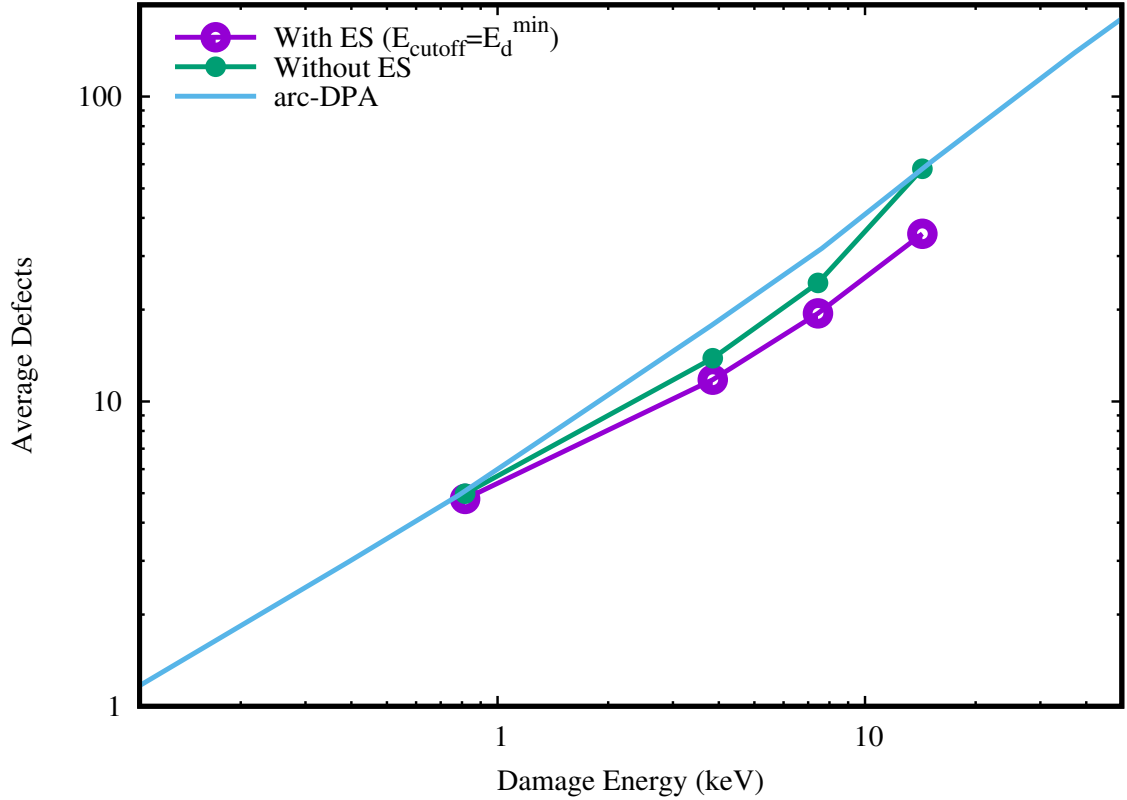}
	\caption{Number of defects in Fe as a function of damage energy in the range 
	1-20 keV, with and without electronic stopping. The number of defects from 
	the arcDPA formula for Fe is also plotted for comparison}
  \label{FeDefects}
\end{figure}

\begin{figure}
  \includegraphics[width=\linewidth]{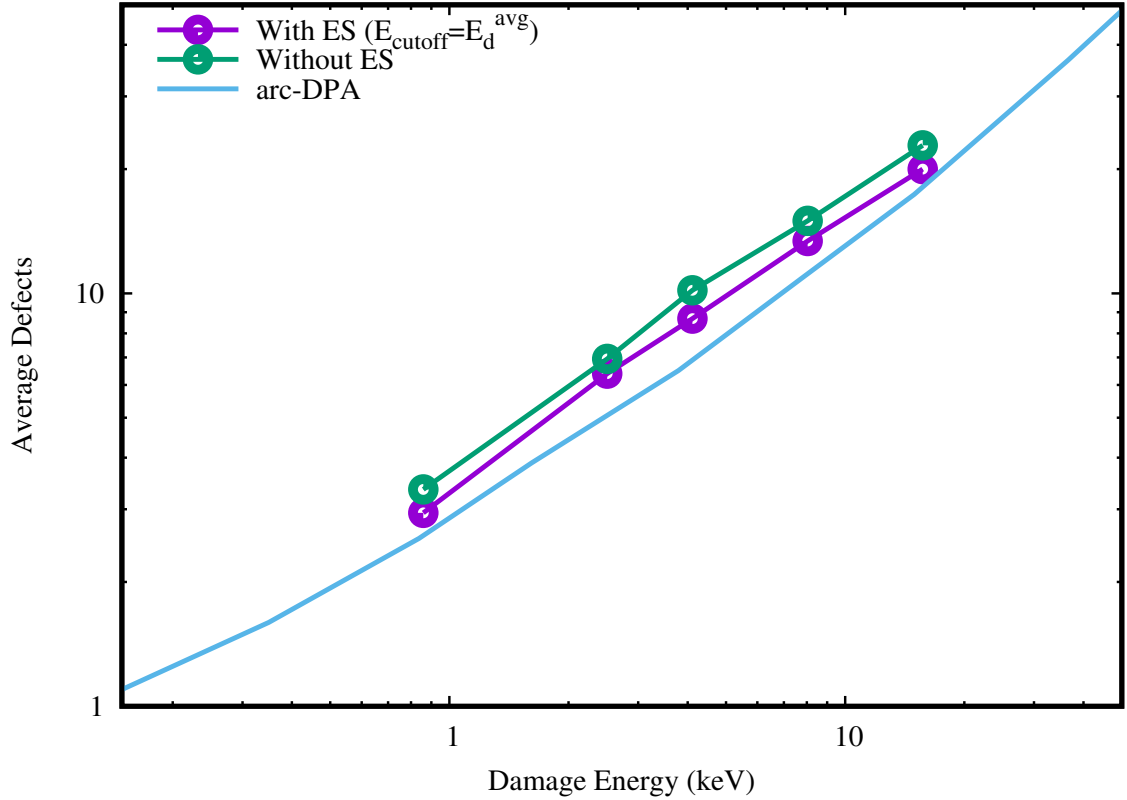}
	\caption{Number of defects in W as a function of damage energy in the range 
	1-20 keV, with and without electronic stopping. The number of defects from 
	the arcDPA formula for W is also plotted for comparison}
  \label{WDefects}
\end{figure}

The direction averaged number of defects with and without ES for both Fe and W 
are shown in Fig.\ref{FeDefects} and Fig.\ref{WDefects} respectively. The 
results are compared with the arcDPA formula \cite{KaiNatureComm2018}. Note 
that at higher PKA energies, ES losses increase and result in lower number of 
defects compared to the case without ES. The losses are higher at higher 
energies resulting in a more pronounced difference in the number of defects 
with and without ES.

\begin{figure}
  \includegraphics[width=\linewidth]{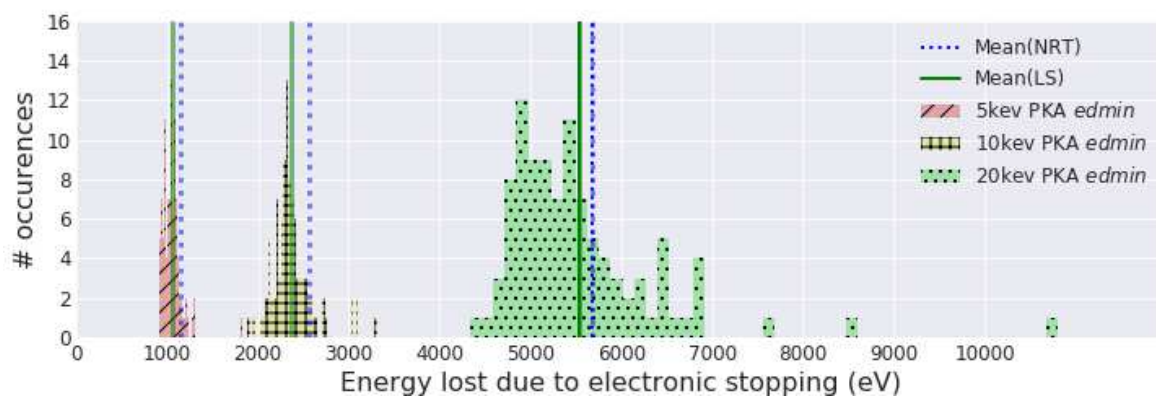}
	\caption{Electronic stopping energy loss at for the 5 keV, 10 keV and 20 keV 
	Fe PKA in Fe using $E_D^{min}$ as the low energy cutoff for ES. The dashed 
	line is the total energy loss calculated from the NRT model and the solid 
	line is the direction averaged total loss from the 100 MD simulations at each 
	of the PKA energies.}
  \label{ESLossCmpFeMin}
\end{figure}

\begin{figure}
  \includegraphics[width=\linewidth]{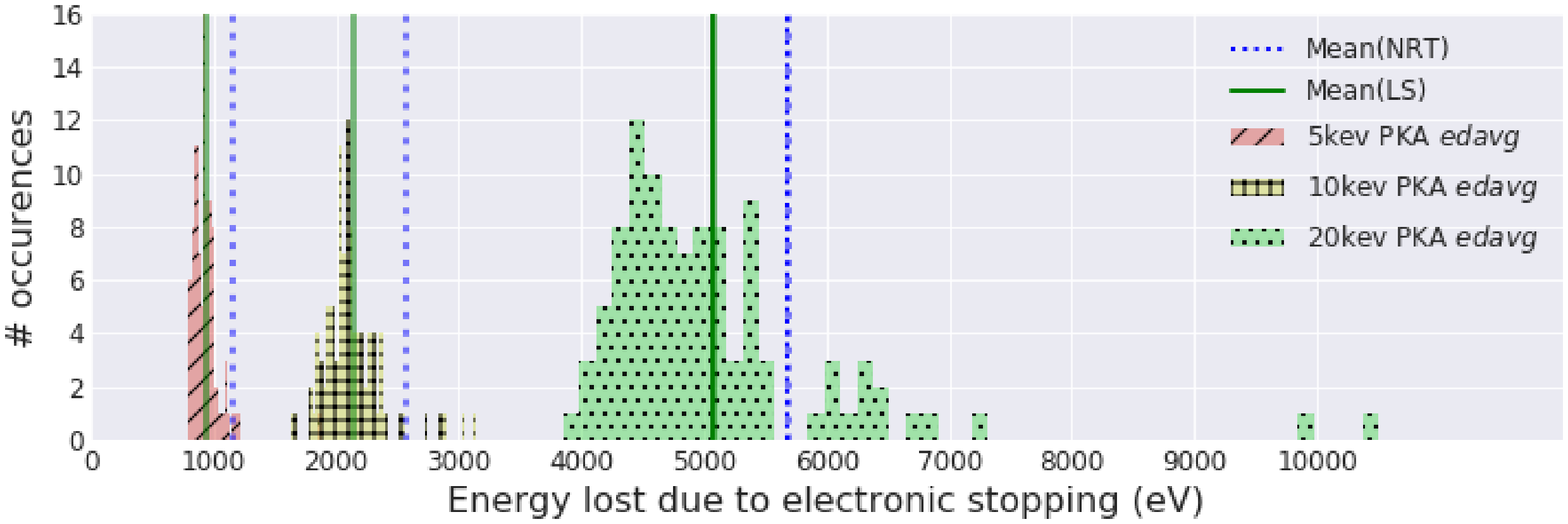}
	\caption{Electronic stopping energy loss at for the 5 keV, 10 keV and 20 keV 
	Fe PKA in Fe using $E_D^{avg}$ as the low energy cutoff for ES. The dashed 
	line is the total energy loss calculated from the NRT model and the solid 
	line is the direction averaged total loss from the 100 MD simulations at each 
	of the PKA energies.}
  \label{ESLossCmpFeAvg}
\end{figure}

Fig.\ref{ESLossCmpFeMin}, Fig.\ref{ESLossCmpFeAvg} and Fig.\ref{ESLossCmpFeMin} 
show the total electronic stopping energy loss at for the 5 keV, 10 keV and 20 
keV Fe PKA in Fe using $E_D^{min}, E_D^{avg}$ and $E_D^{max}$ as the low energy 
cutoff for electronic stopping. It is seen that the choice of $E_D^{min}$ shows 
the best match with the total energy loss calculated by the NRT model. A large 
spread in the total energy loss for PKAs directed in random directions is also 
seen in all the simulations. This variation in the total ES energy loss, is due 
to variations in the collision cascade trajectories for PKAs launched along the 
100 random directions and is a similar effect as the straggling seen in the 
range of energetic atoms in materials. The large spread also has implications 
on the number of defects created because the energy available for damage 
creation will also subsequently have a large spread.

\begin{figure}
  \includegraphics[width=\linewidth]{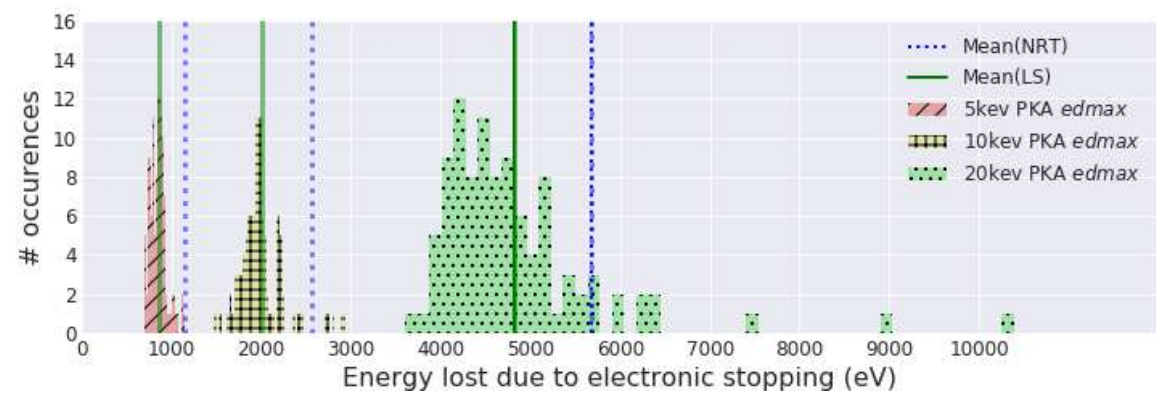}
	\caption{Electronic stopping energy loss at for the 5 keV, 10 keV and 20 keV 
	Fe PKA in Fe using $E_D^{max}$ as the low energy cutoff for ES. The dashed 
	line is the total energy loss calculated from the NRT model and the solid 
	line is the direction averaged total loss from the 100 MD simulations at each 
	of the PKA energies.}
  \label{ESLossCmpFeMax}
\end{figure}

\begin{figure}
  \includegraphics[width=\linewidth]{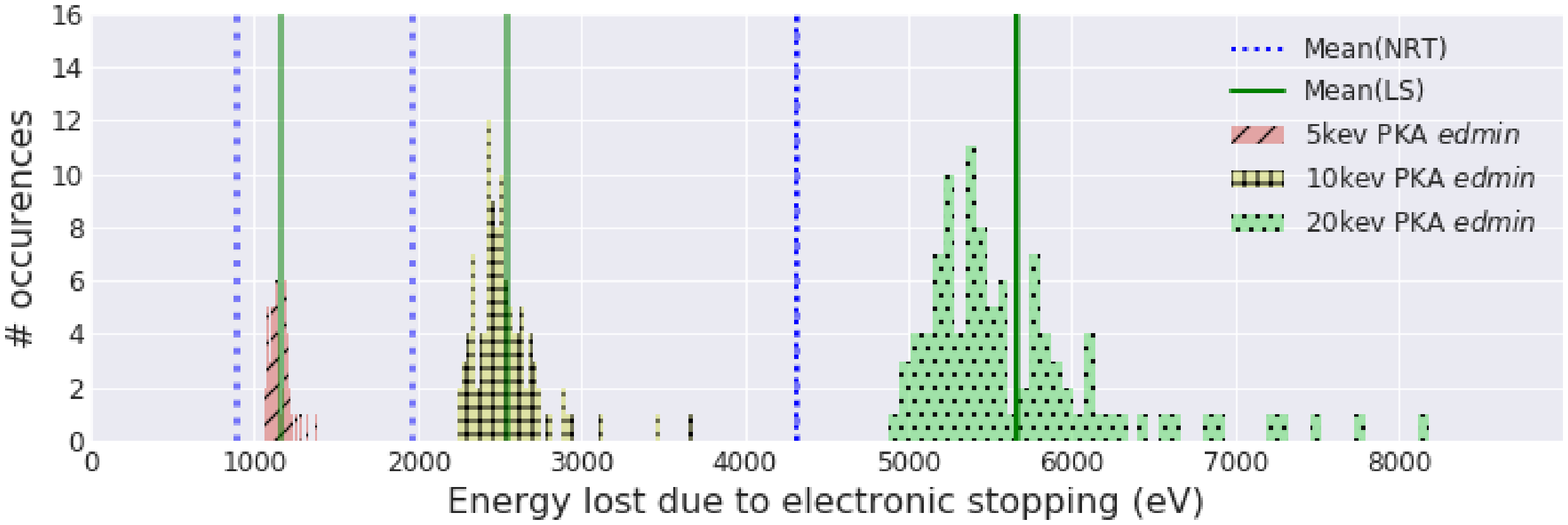}
	\caption{Electronic stopping energy loss at for the 5 keV, 10 keV and 20 keV 
	W PKA in W using $E_D^{min}$ as the low energy cutoff for ES. The dashed 
	line is the total energy loss calculated from the NRT model and the solid 
	line is the direction averaged total loss from the 100 MD simulations at each 
	of the PKA energies.}
  \label{ESLossCmpWMin}
\end{figure}

\begin{figure}
  \includegraphics[width=\linewidth]{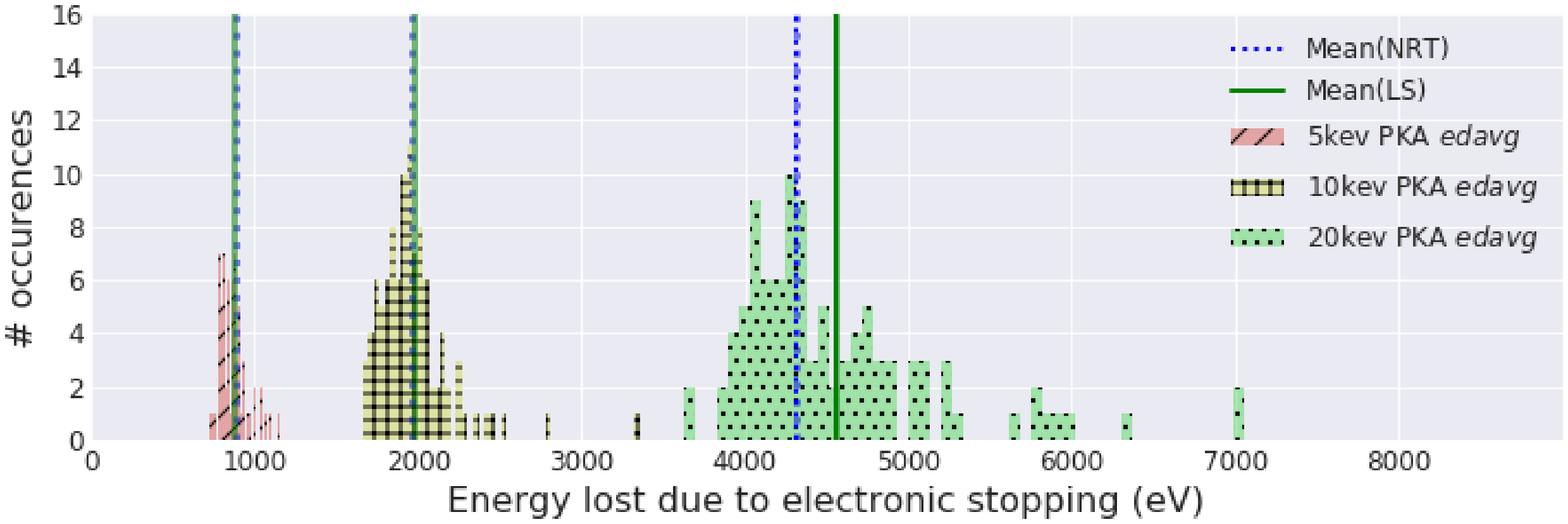}
	\caption{Electronic stopping energy loss at for the 5 keV, 10 keV and 20 keV 
	W PKA in W using $E_D^{avg}$ as the low energy cutoff for ES. The dashed 
	line is the total energy loss calculated from the NRT model and the solid 
	line is the direction averaged total loss from the 100 MD simulations at each 
	of the PKA energies.}
  \label{ESLossCmpWAvg}
  \end{figure}

\begin{figure}
  \includegraphics[width=\linewidth]{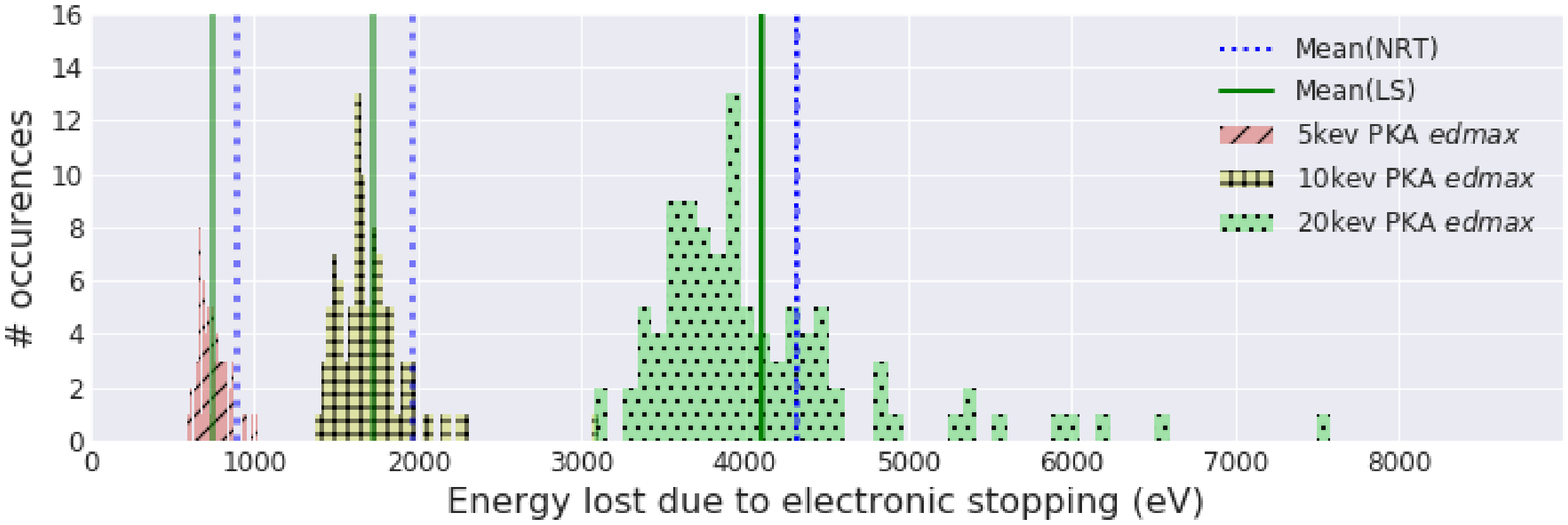}
	\caption{Electronic stopping energy loss at for the 5 keV, 10 keV and 20 keV 
	W PKA in W using $E_D^{max}$ as the low energy cutoff for ES. The dashed 
	line is the total energy loss calculated from the NRT model and the solid 
	line is the direction averaged total loss from the 100 MD simulations at each 
	of the PKA energies.}
  \label{ESLossCmpWMax}
\end{figure}

Similarly, Fig.\ref{ESLossCmpWMin}, Fig.\ref{ESLossCmpWAvg} and 
Fig.\ref{ESLossCmpWMax} show the total electronic stopping energy loss at for 
the 5 keV, 10 keV and 20 keV Fe PKA in W using $E_D^{min}, E_D^{avg}$ and 
$E_D^{max}$ as the low energy cutoff for electronic stopping. The straggling in 
the total energy loss is seen in W too. Regarding the choice of low energy 
cut-off for ES, it is seen that the choice of $E_D^{avg}$ shows the best match 
with the NRT model. This implies that there is no unique way to determine the 
best value of the low energy cutoff for ES, but a value close to $E_D$ is a 
good initial choice.

A discussion on the three methods of including electronic stopping in collision
cascade simulations with LAMMPS might be useful for the reader. A detailed
analysis of the three methods is out of the scope of this document. We mention
our viewpoint on this as follows:

\begin{enumerate}

\item The two temperature model \cite{duffy2006including,rutherford2007effect}
takes care of the local temperature increase of electrons within the collision
cascade and this may have implications on the in-cascade defect clustering and
on defect recombination during the collision cascade. The user has to input
the following parameters:
  \begin{itemize}
  \item electronic specific heat
  \item electronic density
  \item electronic thermal conductivity
  \item friction coefficient due to electron-ion interactions
  \item friction coefficient due to electronic stopping
  \item electronic stopping critical velocity
  \end{itemize}
An electronic sub-system is created within a specified grid in the simulation
domain and the local heating of electrons, the heat transfer between electrons
and ions and the subsequent heat diffusion are all taken into account using
the above inputs. Therefore this is the most comprehensive model taking into
account the maximum physical interactions amongst the three methods and is as
accurate as the models used to input these parameters. Some of these parameters
may not be known for materials of interest and the uncertainity in thee
parameters will contribute to the uncertainity in the results. It is also not
clear how this model treats target materials which have multiple elements
(alloys).

\item The other two methods are same in the sense that both include electronic
stopping as a frictional force. The only difference is that in the existing
method, the user has to input the stopping power as tables and in our method it
is calculated within the code. This difference has the following implications:
  \begin{itemize}
  \item The stopping power tables may be more accurate. However other codes
  have to be run in order to create the tables if not available to the user.
  \item Using stopping power tables allows calculation of electronic stopping
  in the case of alloys in principle.
  \item The method developed in this work currently assumes a single element
  target. The friction coefficient and the expected energy loss due to
  electronic stopping as given by the NRT model need to be calculated only once
  at beginning of the simulation.
  \end{itemize}

\end{enumerate}
A detailed study of the three methods and their effects on number of defects
and on defect clutering needs to be carried out. This is out of scope of this
paper which aims to describe our implementation of electronic stopping in
LAMMPS.

\section{Conclusions}
Electronic stopping losses, which decreases the available energy to create 
defects in collision cascades, has been included in the open source Large-scale 
Atomic Molecular Massively Parallel  Simulator (LAMMPS). This is done by 
implementing a new "fix" in LAMMPS called \begin{verbatim} fix_esfriction 
\end{verbatim} which adds a friction force to all atoms moving with a velocity 
greater than a specified cut-off velocity. The co-efficient of the frictional 
force is obtained from the Lindhard-Scharff model. The code is validated by 
comparing the total ES loss from LAMMPS simulations with the total ES from NRT 
model. It is shown that ES affects the number of defects created, and at high 
energies of PKA, the effect of ES on the number of defects formed is more 
pronounced as expected. The total electronic stopping losses along different
directions also show a large deviation due to straggling along the random
directions of launch of the PKA and as expected, larger the PKA energy, more
is the straggling. An initial choice of the low energy cut-off for ES can
be the displacement energy of the atom and finer adjustments to the chosen
value can be made by adjusting it so that the average total energy loss due to
ES matches with that calculated by the NRT model.

\noindent
{\bf References}\\

\end{document}